\documentclass[prc,twocolumn,superscriptaddress,showpacs,aps]{revtex4}
\usepackage{graphicx}
\begin{document}


\title{ Measurement of the $\vec{\gamma} p \to K^+ \Lambda$ Reaction 
	at Backward Angles }

\author{K.~Hicks}
  \affiliation{Department of Physics and Astronomy, Ohio University, Athens, Ohio 45701, USA}
\author{T.~Mibe}
  \affiliation{Department of Physics and Astronomy, Ohio University, Athens, Ohio 45701, USA}
\author{M.~Sumihama}
  \affiliation{Research Center for Nuclear Physics, Osaka University, Ibaraki, Osaka 567-0047, Japan}
\author{D.S.~Ahn}
  \affiliation{Research Center for Nuclear Physics, Osaka University, Ibaraki, Osaka 567-0047, Japan}
  \affiliation{Department of Physics, Pusan National University, Busan 609-735, Korea}
\author{J.K.~Ahn}
  \affiliation{Department of Physics, Pusan National University, Busan 609-735, Korea}
\author{H.~Akimune}
  \affiliation{Department of Physics, Konan University, Kobe, Hyogo 658-8501, Japan}
\author{Y.~Asano}
  \affiliation{Synchrotron Radiation Research Center, Japan Atomic Energy Research Institute, Mikazuki, Hyogo 679-5198, Japan}
\author{W.C.~Chang}
  \affiliation{Institute of Physics, Academia Sinica, Taipei 11529, Taiwan}
\author{S.~Dat\'e}
  \affiliation{Japan Synchrotron Radiation Research Institute, Mikazuki, Hyogo 679-5198, Japan}
\author{H.~Ejiri}
  \affiliation{Research Center for Nuclear Physics, Osaka University, Ibaraki, Osaka 567-0047, Japan}
  \affiliation{Japan Synchrotron Radiation Research Institute, Mikazuki, Hyogo 679-5198, Japan}
\author{S.~Fukui}
  \affiliation{Department of Physics and Astrophysics, Nagoya University, Nagoya, Aichi 464-8602, Japan}
\author{H.~Fujimura}
  \affiliation{Department of Physics, Kyoto University, Kyoto 606-8502, Japan} 
  \affiliation{School of Physics, Seoul National University, Seoul, 151-747, Korea}
\author{M.~Fujiwara}
  \affiliation{Research Center for Nuclear Physics, Osaka University, Ibaraki, Osaka 567-0047, Japan}
  \affiliation{Kansai Photon Science Institute, Japan Atomic Energy Agency, Kizu, Kyoto, 619-0215, Japan}
\author{S.~Hasegawa}
  \affiliation{Research Center for Nuclear Physics, Osaka University, Ibaraki, Osaka 567-0047, Japan}
\author{A.~Hosaka}
  \affiliation{Research Center for Nuclear Physics, Osaka University, Ibaraki, Osaka 567-0047, Japan}
\author{T.~Hotta}
  \affiliation{Research Center for Nuclear Physics, Osaka University, Ibaraki, Osaka 567-0047, Japan}
\author{K.~Imai}
  \affiliation{Department of Physics, Kyoto University, Kyoto 606-8502, Japan} 
\author{T.~Ishikawa}
  \affiliation{Laboratory of Nuclear Science, Tohoku University, Sendai, Miyagi 982-0826, Japan}
  \affiliation{Department of Physics, Kyoto University, Kyoto 606-8502, Japan} 
\author{T.~Iwata}
  \affiliation{Department of Physics, Yamagata University, Yamagata 990-8560, Japan}
\author{B. Juli\'a-D\'iaz}
  \affiliation{Departament de Estructura i Constituents de la Materia, Diagonal 647, 08028 Barcelona, Spain}
\author{Y.~Kato}
  \affiliation{Research Center for Nuclear Physics, Osaka University, Ibaraki, Osaka 567-0047, Japan}
\author{H.~Kawai}
  \affiliation{Department of Physics, Chiba University, Chiba 263-8522, Japan}
\author{Z.Y.~Kim}
  \affiliation{School of Physics, Seoul National University, Seoul, 151-747, Korea}
\author{K.~Kino}
  \altaffiliation[Present address: ]{Center for Nuclear Study, University of Tokyo,7-3-1 Hongo, Bunkyo, Tokyo 113-0033, Japan}
  \affiliation{Research Center for Nuclear Physics, Osaka University, Ibaraki, Osaka 567-0047, Japan}
\author{H.~Kohri}
  \affiliation{Research Center for Nuclear Physics, Osaka University, Ibaraki, Osaka 567-0047, Japan}
\author{N.~Kumagai}
  \affiliation{Japan Synchrotron Radiation Research Institute, Mikazuki, Hyogo 679-5198, Japan}
\author{T.-S.~H.~Lee}
  \affiliation{Physics Division, Argonne National Laboratory, Argonne, Illinois 60439, USA}
\author{S.~Makino}
  \affiliation{Wakayama Medical University, Wakayama, 641-8509, Japan}
\author{T.~Matsuda}
  \affiliation{Department of Applied Physics, Miyazaki University, Miyazaki 889-2192, Japan}
\author{T.~Matsumura}
  \affiliation{Department of Applied Physics, National Defense Academy, Yokosuka 239-8686, Japan}
\author{N.~Matsuoka}
  \affiliation{Research Center for Nuclear Physics, Osaka University, Ibaraki, Osaka 567-0047, Japan}
\author{M.~Miyabe}
  \affiliation{Department of Physics, Kyoto University, Kyoto 606-8502, Japan} 
\author{Y.~Miyachi}
  \altaffiliation[Present address: ]{Department of Physics, Tokyo Institute of Technology, Tokyo 152-8551, Japan}
  \affiliation{Department of Physics and Astrophysics, Nagoya University, Nagoya, Aichi 464-8602, Japan}
\author{M.~Morita}
  \affiliation{Research Center for Nuclear Physics, Osaka University, Ibaraki, Osaka 567-0047, Japan}
\author{N.~Muramatsu}
  \affiliation{Research Center for Nuclear Physics, Osaka University, Ibaraki, Osaka 567-0047, Japan}
\author{T.~Nakano}
  \affiliation{Research Center for Nuclear Physics, Osaka University, Ibaraki, Osaka 567-0047, Japan}
\author{M.~Niiyama}
  \affiliation{Department of Physics, Kyoto University, Kyoto 606-8502, Japan} 
\author{M.~Nomachi}
  \affiliation{Department of Physics, Osaka University, Toyonaka, Osaka 560-0043, Japan}
\author{Y.~Ohashi}
  \affiliation{Japan Synchrotron Radiation Research Institute, Mikazuki, Hyogo 679-5198, Japan}
\author{T.~Ooba}
  \affiliation{Department of Physics, Chiba University, Chiba 263-8522, Japan}
\author{H.~Ohkuma}
  \affiliation{Japan Synchrotron Radiation Research Institute, Mikazuki, Hyogo 679-5198, Japan}
\author{D.S.~Oshuev}
  \affiliation{Institute of Physics, Academia Sinica, Taipei 11529, Taiwan}
\author{S.~Ozaki}
  \affiliation{Research Center for Nuclear Physics, Osaka University, Ibaraki, Osaka 567-0047, Japan}
\author{C.~Rangacharyulu}
  \affiliation{Department of Physics and Engineering Physics, University of Saskatchewan, Saskatoon, Saskatchewan, Canada, S7N 5E2} 
\author{A.~Sakaguchi}
  \affiliation{Department of Physics, Osaka University, Toyonaka, Osaka 560-0043, Japan}
\author{T.~Sasaki}
  \affiliation{Department of Physics, Kyoto University, Kyoto 606-8502, Japan} 
\author{P.M.~Shagin}
  \altaffiliation[Present address: ]{ Department of Physics and Astronomy, Rice University, 6100 Main St. Houston MS 108, TX 77005-1892, USA}
  \affiliation{Research Center for Nuclear Physics, Osaka University, Ibaraki, Osaka 567-0047, Japan}
\author{Y.~Shiino}
  \affiliation{Department of Physics, Chiba University, Chiba 263-8522, Japan}
\author{A.~Shimizu}
  \affiliation{Research Center for Nuclear Physics, Osaka University, Ibaraki, Osaka 567-0047, Japan}
\author{H.~Shimizu}
  \affiliation{Laboratory of Nuclear Science, Tohoku University, Sendai, Miyagi 982-0826, Japan}
\author{Y.~Sugaya}
  \affiliation{Department of Physics, Osaka University, Toyonaka, Osaka 560-0043, Japan}
  \affiliation{Advanced Science Research Center, Japan Atomic Energy Research Institute, Tokai, Ibaraki 319-1195, Japan}
\author{Y.~Toi}
  \affiliation{Department of Applied Physics, Miyazaki University, Miyazaki 889-2192, Japan}
\author{H.~Toyokawa}
  \affiliation{Japan Synchrotron Radiation Research Institute, Mikazuki, Hyogo 679-5198, Japan}
\author{A.~Wakai}
  \altaffiliation[Present address: ]{Akita Research Institute of Brain and Blood Vessels, Akita 010-0874, Japan}
  \affiliation{Center for Integrated Research in Science and Engineering, Nagoya University, Nagoya, Aichi 464-8603, Japan}
\author{C.W.~Wang}
  \affiliation{Institute of Physics, Academia Sinica, Taipei 11529, Taiwan}
\author{S.C.~Wang}
  \altaffiliation[Present address: ]{Institute of Statistical Science, Academia Sinica, Nankang, 115 Taipei, Taiwan}
  \affiliation{Institute of Physics, Academia Sinica, Taipei 11529, Taiwan}
\author{K.~Yonehara}
  \altaffiliation[Present address: ]{Illinois Institute of Technology, Chicago, Illinois 60616, USA}
  \affiliation{Department of Physics, Konan University, Kobe, Hyogo 658-8501, Japan}
\author{T.~Yorita}
  \affiliation{Japan Synchrotron Radiation Research Institute, Mikazuki, Hyogo 679-5198, Japan}
\author{M.~Yoshimura}
  \affiliation{Institute for Protein Research, Osaka University, Suita, Osaka 565-0871, Japan}
\author{M.~Yosoi}
  \affiliation{Research Center for Nuclear Physics, Osaka University, Ibaraki, Osaka 567-0047, Japan}
  \affiliation{Department of Physics, Kyoto University, Kyoto 606-8502, Japan} 
\author{R.G.T.~Zegers}
  \affiliation{National Superconducting Cyclotron Laboratory, Michigan State University, East lansing, MI 48824-1321, USA}

\collaboration{The LEPS collaboration}
\noaffiliation

\date{\today}

\begin{abstract}
Cross sections for the $\gamma p \to K^+ \Lambda$ have been measured 
at backward angles using linearly polarized photons in the range 
1.50 to 2.37 GeV.  In addition, the beam asymmetry for this reaction 
has been measured for the first time at backward angles. 
The $\Lambda$ was detected at forward angles in the LEPS spectrometer
via its decay to $p\pi^-$ and the $K^+$ was inferred using the technique 
of missing mass.  These measurements, corresponding to kaons at far 
backward angles in the center-of-mass frame, complement similar CLAS 
data at other angles.  Comparison with theoretical models shows that 
the reactions in these kinematics provide further opportunities to 
investigate the reaction mechanisms of hadron dynamics.
\end{abstract}

\pacs{13.60.Le, 25.20.Lj}

\maketitle

Measurements of the photoproduction of the $K^+\Lambda$ final 
state with high statistics have become possible in the past 
decade due to high-flux photon beams in the GeV range. Recently, 
the CLAS collaboration published a compendium of data 
for this reaction over a wide range of 
angles and photon energies \cite{bradford}.  
One motivation for more complete $K^+\Lambda$ data is to study the  
details of $N^*$ resonances that were predicted to couple weakly 
to pion decay and more strongly to kaon decay.  
Some $N^*$ resonances that were predicted in quark models 
\cite{capstick}, but were not seen in partial wave analysis 
of pion scattering data, might be seen in kaon production.
The CLAS data do not show definitive evidence 
for new $N^*$ resonances, but do exhibit a few broad 
energy-dependent structures in the differential cross sections, 
suggesting that there is a more complicated mechanism
contributing to the $\gamma p \to K\Lambda$ reaction.
The present results extend the existing data for 
$K^+\Lambda$ photoproduction to far-backward kaon angles.

Theoretical progress for $K^+\Lambda$ photoproduction has been 
published recently \cite{bruno} showing that coupled channel 
effects can no longer be ignored.  This approach, the 
dynamical coupled-channels (DCC) formalism, includes a proper 
treatment of off-shell effects. The most important multistep 
transition, $\gamma N \to \pi N \to K \Lambda$, has a comparable 
cross section to direct production, $\gamma N \to K \Lambda$ 
\cite{tabakin}.  In this case, the DCC formalism is 
necessary for a correct interpretation of $K^+\Lambda$ 
photoproduction data.

The CLAS detector acceptance does not allow measurements at 
either very forward or very backward kaon angles.  
Recently, the LEPS collaboration published cross sections 
for forward angles\cite{sumihama}, showing good 
agreement with the CLAS data where the two data sets overlap.
Here, cross sections for backward angles of this reaction are 
measured using a different experimental technique, 
where the $\Lambda$ is reconstructed from its decay products 
which are detected in the LEPS spectrometer.  The final state 
kaon, which goes backward in the center-of-mass frame, is 
measured through the missing mass technique.

\begin{figure}[tb] 
\includegraphics[width=8.5cm]{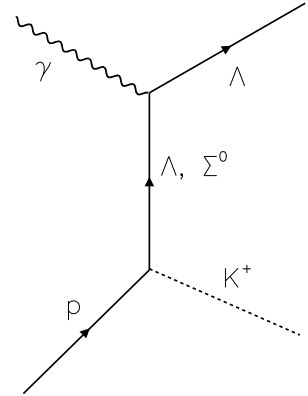} 
\caption{ Diagram of the $u$-channel for the reaction 
$\gamma p \to K^+\Lambda$ which contributes at backward kaon angles.
} 
\label{fig:diag} 
\end{figure} 

In the kinematics that LEPS can access, we can study certain reaction 
dynamics selectively.  For instance, in meson production at forward 
angles, $t$-channel diagrams dominate, and the nature of the 
exchange particle can be studied (see Ref. \cite{mibe} for an 
example in $\phi$ photoproduction).  In contrast, the far backward 
angles in meson photoproduction are associated with $u$-channel diagrams.
In this paper, we will examine whether the data exhibit 
any $u$-channel signature in the cross sections.

At low energies, the $u$-channel amplitude is expected to be 
dominated by diagrams where a hyperon, such as a ground state 
$\Lambda$ or $\Sigma^0$, is exchanged.  These diagrams are shown 
in Fig. \ref{fig:diag}.  The coupling constants for the hadronic 
verticies in Fig. \ref{fig:diag} are inferred from the $t$-channel 
in kaon photoproduction data, which are determined either by 
theoretical models (e.g. SU(3) symmetry) or by an independent fit in 
analyses such as the DCC formalism for $\pi N \to KY$ reactions.
The exchange baryon is neutral, and so the electromagnetic vertex 
is dominated by the M1 multipole which acts to flip the spin of 
the $\Lambda$ or $\Sigma^0$.  The magnetic moments of the $\Lambda$ 
and $\Sigma$ hyperons are known, so this vertex has little ambiguity.
There is not much freedom in calculations of the diagram shown in 
Fig. \ref{fig:diag}.  

Of course, we expect that there are further dynamical processes 
which may contribute to the reaction at this kinematical region. 
In the low energy regime, for instance, coupled channel effects and 
nucleon resonances may be important as emphasized in Ref. \cite{bruno}.  
The Regge model has also been used in the study of reactions in 
this energy region as well as with higher energy photons 
\cite{guidal,mart}.  In this description, one can study the nature 
of baryon trajectories exchanged in the $u$-channel.
The purpose of this paper is to provide the cross sections and 
spin asymmetries at far backward angles and discuss reaction 
mechanisms which may be relevant to $u$-channel dynamics.

Direct detection of the $\Lambda$ 
provides additional information.  At LEPS, the incident 
photon is highly polarized, and so 
the reaction plane of the $K^+\Lambda$ can be compared with the plane 
of the photon polarization, giving new information on the 
reaction mechanism.  The beam asymmetry for kaon photoproduction 
has been measured at forward angles by LEPS \cite{zegers}, and  
is measured here in $u$-channel kinematics for the first time.


The experimental data were taken using the LEPS (laser electron 
photons at SPring-8) detector in Japan \cite{nakano}.  Ultraviolet 
light from an Ar laser was linearly polarized and directed onto the 
8 GeV stored electron beam.  Backward Compton scattering produced 
a narrow beam of photons up to 2.4 GeV.  The struck electron 
was detected in a tagging spectrometer, giving the energy of 
individual photons in the range 1.5-2.4 GeV.  The linear polarization 
of the photons is calculated for Compton scattering and was 
typically 97\% at the maximum energy.  The photon beam was 
incident on a 16 cm long liquid hydrogen target.  Details of 
the geometry are given elsewhere \cite{sumihama}.

The LEPS spectrometer consists of a wide-gap dipole magnet, with 
charged-particle tracking detectors both before and after the 
magnet. An array of scintillator bars were placed 4 meters 
downstream of the target, and along with a start counter (SC) 
scintillator 5 cm downstream of the target, provided a time-of-flight 
(TOF) measurement.  The trigger was a coincidence between the 
tagger, the SC, and the TOF wall.  Electron-positron pairs 
were vetoed by an aerogel cerenkov detector just after the SC.

The total number of photons 
on target was $1.18 \times 10^{12}$, after correcting for the 
transmission factor (for material between the beam production point 
and the target) and tagger inefficiencies.  


\begin{figure}[tb] 
\includegraphics[width=8.5cm]{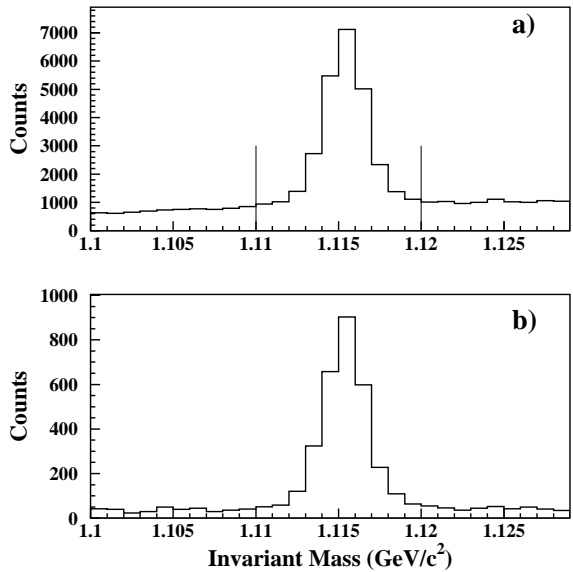} 
\caption{ Invariant mass of the $p\pi^-$ events, showing a 
peak for the $\Lambda$ riding on top of a smooth background:
a) raw data with a reconstructed vertex from the target, 
b) with an additional requirement that the missing mass is a kaon.
} 
\label{fig:lam} 
\end{figure} 

Using momentum and TOF, the mass of each detected 
particle was measured, giving particle identification (PID). 
Events with a proton and a $\pi^-$, each having a measured 
mass within $3\sigma$ (where $\sigma$ is the momentum dependent 
mass resolution) of its known mass were kept for further 
analysis. The point of closest approach between these two 
tracks was calculated, and this vertex position was required 
to be within the target or downstream of the target.
Because of the long lifetime of the $\Lambda$, 
the vertex position can be downstream of the target.  
A cut on vertex position before the SC was required in the analysis.  
Empty target runs showed that the contribution 
of the target windows and the SC was less than about 4\%. 
Monte Carlo simulations showed good agreement with the 
distribution of vertex position for events with 
$\Lambda \to p\pi^-$ decay in the experiment.

In addition to mass cuts, several additional requirements 
are used to ensure good PID.  For example, when the track 
is extrapolated to the TOF wall, the position obtained from 
timing measurements on either end of the hit TOF bar must 
be within 8 cm of the expected position.  The same
PID requirements as described in Ref. \cite{sumihama} are 
used in the present analysis.

Fig. \ref{fig:lam} shows the invariant mass of the $p\pi^-$ 
pair for: (a) all events with good PID and the above vertex cut 
and (b) for those events where the missing mass is consistent 
with the kaon mass.  The smooth background 
in Fig. \ref{fig:lam}a likely comes from reactions like 
$\gamma p \to p \pi^+\pi^-$ where the $\pi^+$ is not detected. 
The spectrum in Fig. \ref{fig:lam}b has much less background because 
the missing particle is now required to have the mass of a $K^+$.
The technique of sideband subtraction can be used to remove 
the remaining background.  Let the $\Lambda$ region be given by 
a cut on the mass from 1.110 to 1.120 GeV/c$^2$ (shown by the 
vertical lines in Fig. \ref{fig:lam}).  The left and right 
sideband regions on either side of the peak (of equal width) 
were analyzed in the same way as the $\Lambda$ 
region and subtracted from the final results.

\begin{figure}[htb] 
\includegraphics[width=9.0cm]{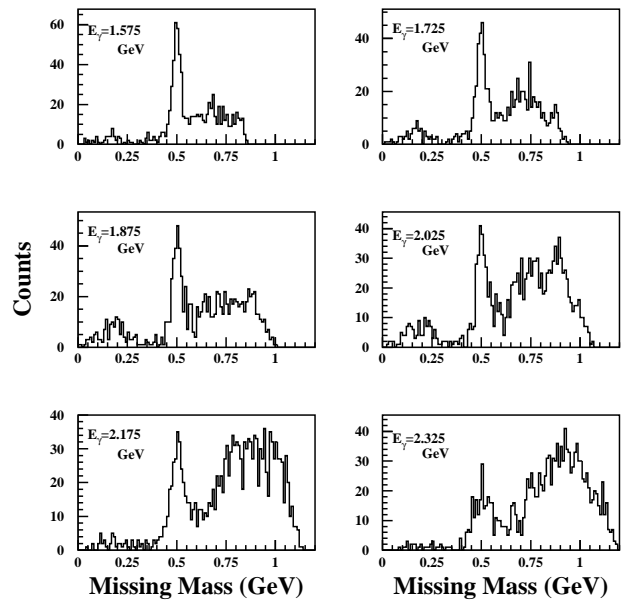} 
\caption{ 
Missing mass of the $\gamma p \to \Lambda X$ reaction 
for the angular bin $0.90 < \cos\theta_{CM} < 1.0$ where 
$\theta_{CM}$ is the center-of-mass angle of the $\Lambda$.  
Each figure is for the photon energy shown by the label.
} 
\label{fig:fits} 
\end{figure} 

The missing mass for the reaction $\gamma p \to \Lambda X$ is 
shown in Fig. \ref{fig:fits} for six equal bins in the photon energy.  
The photon energy bins are 150 MeV wide, in six equal steps 
from 1.5 to 2.4 GeV.  The number of photons in each energy 
bin was measured by the tagger, corrected for the inefficiencies 
of each tagger element. The plots in Fig. \ref{fig:fits} are 
shown in order of increasing energy bin, from lowest (top left) 
to highest (lower right).  

The data in Fig. \ref{fig:fits} span the range 
$0.9 < \cos\theta_{CM} < 1.0$ 
where $\theta_{CM}$ is the center-of-mass (CM)  polar angle of the 
$\Lambda$ momentum vector.  These spectra have not yet been 
sideband subtracted, which mostly removes counts below a mass 
of 0.4 GeV/c$^2$.  A clear peak is seen in the missing 
mass spectra at the value of the known $K^+$ mass. The 
strength at higher mass corresponds to a combination of 
$K\Sigma$ (followed by $\Sigma \to \Lambda\gamma$), $K^*\Lambda$ 
and $KY^*$ photoproduction.  

The $K^+$ peak appears to decrease rapidly with increasing 
photon energy.  These spectra are not corrected 
for the the acceptance for $\Lambda$ detection in the LEPS 
spectrometer, which is weakly dependent on photon energy. 
Data in a second angular bin, $0.8 < \cos\theta_{CM} < 0.9$, 
are of similar quality and statistics.

The LEPS acceptance was calculated based on Monte Carlo 
simulations for $K^+\Lambda$ production uniformly distributed 
in energy and center-of-mass angle.  More realistic distributions 
are possible, and studies with a phenomenological energy-dependent 
event generator showed that the systematic uncertainty 
of the acceptance is on the order of 4\% or less.  The 
simulations were carried out using the GEANT software \cite{geant} 
with input for the detector geometry and resolutions.
A realistic Compton scattered photon beam distribution was used. 
The simulated peak widths for the invariant mass ($\Lambda$) 
and missing mass ($K^+$) are in good agreement with those 
shown in Figs. \ref{fig:lam} and \ref{fig:fits}.

In addition to the $K^+\Lambda$ final state, the three reactions 
at higher missing mass mentioned above ($K\Sigma$, $K^*\Lambda$, $KY^*$) 
were simulated along with a general 3-body phase space for 
the $K\pi\Lambda$ final state.  Missing mass spectra from all 
of the simulations were used as input to a template fit of the 
experimental data, where each template spectrum was multiplied 
by an overall factor to minimize the reduced $\chi^2$, which was 
typically in the range of 1-2.  The number of counts in the 
$K^+$ peak was extracted from the template fits. The $K^+$ fit 
is affected primarily by background from the $K^+\Sigma^0$ reaction, 
which was constrained largely by the events at masses about 
0.1 GeV/c$^2$ higher than the $K^+$ peak.  Simulations show 
that the other reactions, including 3-body phase space, have 
significant strength only at higher missing mass than for $K\Sigma$. 

The systematic uncertainty of the fitting procedure was estimated by 
doing fits with gaussians for the background of 
$K\Sigma$ production and of the 3-body final states at higher missing 
mass. Comparison of the Gaussian and template fit results gives a 
mean systematic uncertainty of 5\%. Other systematic uncertaintes 
due to target thickness, photon flux, and event selection cuts 
added in quadrature contributes 4\%. The overall systematic 
uncertainty, including that of the Monte Carlo simulations, is 7.5\%.

\begin{figure}[htb] 
\includegraphics[width=8.5cm]{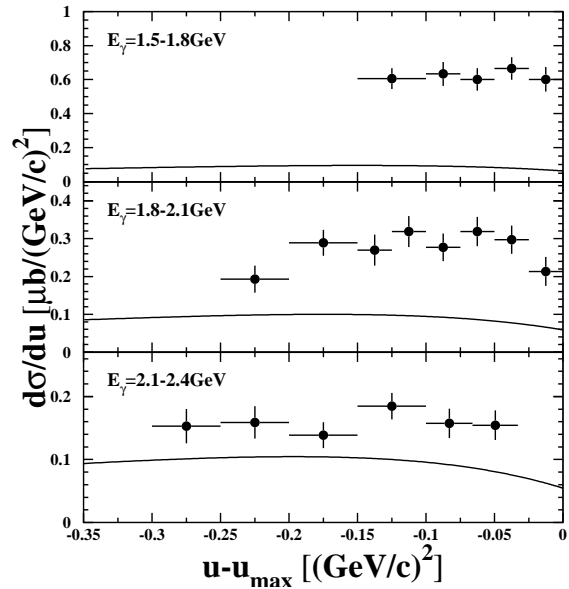} 
\caption{ (color online)
Differential cross sections as a function of the Mandelstam variable 
$u$ for the given photon beam energy.  The curves are calculations 
of the contribution from the diagram in Fig. 1 only, using exchange 
of $\Lambda$, $\Sigma^0$ and $\Sigma^*$ hyperons.
} 
\label{fig:dsdu} 
\end{figure}

For three photon energy ranges, see Fig. \ref{fig:dsdu}, 
the cross section was binned for several values of 
$u = (p_\gamma - p_\Lambda)^2$.  The maximum 
value, $u_{max}$ occurs when the $\Lambda$ goes forward at $0^\circ$ 
from the photon direction.  The cross sections in Fig. \ref{fig:dsdu} 
are shown as a function of $u-u_{max}$.  
Theoretical calculations \cite{hosaka} for the $u$-channel only, 
shown by the solid line, are far below the data and suggest that 
the diagram of Fig. \ref{fig:diag} is not dominant.  These calculations 
include exchange baryons of $\Lambda$, $\Sigma^0$, and $\Sigma(1385)$ 
along with a form factor with a cut-off mass of about 0.9 GeV (the 
theoretical values increase for higher cut-off mass).
It appears that $s$-channel diagrams still contribute strongly 
to the cross section even at these far backward kaon angles, at 
least at the lower photon energies.

At photon energies above 4.3 GeV, $K^+\Lambda$ data clearly show 
a rise for $u$ between $-0.2$ to $-0.7$ (GeV/c)$^2$ \cite{anderson},
but the present data do not exhibit this $u$-channel signature.
The fact that the cross sections are nearly constant as a function 
of $u-u_{max}$ can be interpreted as additional evidence for the 
lack of dominance by the $u$-channel diagram of Fig. \ref{fig:diag} 
for backward-angle $K^+\Lambda$ data at photon energies below about 
3 GeV.

\begin{figure}[htb] 
\includegraphics[width=8.5cm]{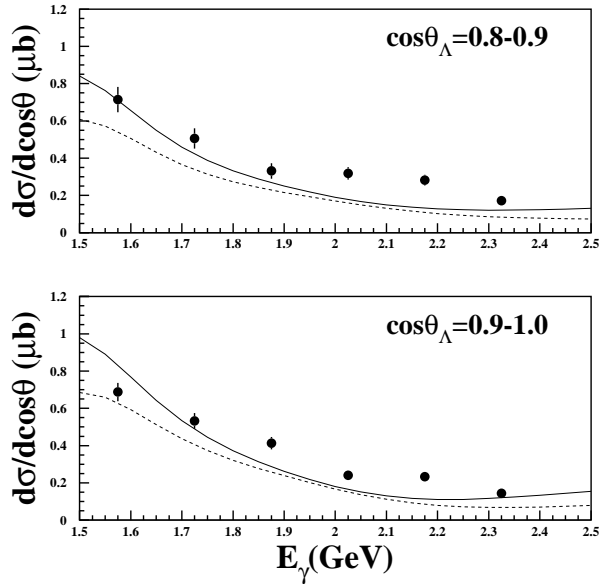} 
\caption{ (color online)
Differential cross sections at the angle shown as a function of 
photon beam energy.  Calculations in the model of Ref. \cite{bruno} 
are shown without (dashed) and with (solid line) coupled channels.
} 
\label{fig:xsec} 
\end{figure}

Differential cross sections as a function of six photon energy bins, 
for each of two $\Lambda$ angle bins (see above), 
are plotted in Fig. \ref{fig:xsec}.  Here, the cross sections 
are presented in the same format as for the CLAS data \cite{bradford}. 
The angle-dependent acceptance ranged from about 2-3\% in 
the lowest energy bin up to 6-8\% at the highest energy bin.  

Since the $u$-channel diagram alone can not explain the present 
data, 
theoretical curves from Ref. \cite{bruno} are shown in 
Fig. \ref{fig:xsec}, converted to the present units. The 
solid curve is for the full dynamical coupled-channels (DCC) 
model, whereas the dashed curves do not include DCC effects
(see Fig. 14 of Ref. \cite{bruno}). The present data are closer to 
the full DCC calculation, except for one point at the lowest photon 
energy ($E_\gamma=1.575$) and most backward kaon angle.  
The top plot, for $\cos\theta_{CM}=0.85$ agrees within error bars 
with the CLAS data, which is conveniently tabulated in Ref. 
\cite{bradford} (but not shown in Fig. \ref{fig:xsec}a).  
The lower plot, for $\cos\theta_{CM}=0.95$, 
goes beyond the angular region covered by CLAS. 
In general the data follows the energy and angle 
dependence predicted by the DCC model, but are still 
significantly different in the range of $E\gamma=2.0$-2.2 GeV. 

Alternatively, we have performed a simple estimation using 
the Regge model, where we fit the energy dependence by an 
exponential function, 
$$ \frac{d \sigma}{du} \propto s^{-2(\alpha(0)-1)} $$
where $\alpha(0)$ is the intercept for the hyperon 
trajectory exchanged in the $u$-channel.  By choosing
$\alpha(0)=-0.84$, the energy dependence can be fit and 
extrapolated to the previous data at 4.3 GeV \cite{anderson}.
This value of $\alpha(0)$ is not too far from the value $-0.68$ 
extrapolated from the mass dependence of the $\Lambda^*$ 
resonances \cite{guidal}.

\begin{figure}[htb] 
\includegraphics[width=8.5cm]{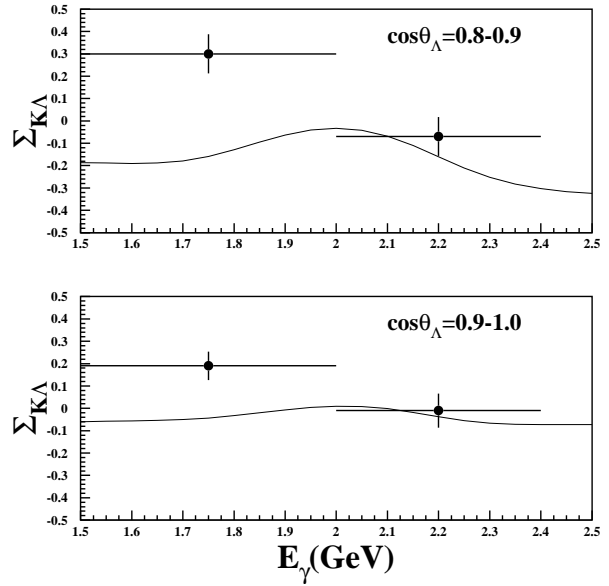} 
\caption{ 
Beam asymmetry at the angles shown as a function of 
photon beam energy.  The points have been averaged over a range 
of photon energy as described in the text.  Calculations in the 
model of Ref. \cite{bruno} are shown for comparison.
} 
\label{fig:asym} 
\end{figure} 

Fig. \ref{fig:asym} shows the beam asymmetry, which was measured 
by binning the $K^+\Lambda$ data as a function of $\phi$, which 
is defined as the azimuthal angle of the $\Lambda$ as measured 
from the linear polarization plane of the photon beam.  In order 
to gain sufficient statistics for the $\phi$-fit, only two 
bins in photon energy were used, one from 1.5-2.0 GeV, and a 
second one from 2.0-2.4 GeV.  We follow the same procedure 
to extract the beam asymmetry as described previously \cite{zegers}. 
The results are a positive beam asymmetry, $\Sigma_{K\Lambda}$, 
in the first energy bin, and a slightly negative polarzation 
at higher photon energy. The positive asymmetry means that 
more $K^+\Lambda$ reactions are produced perpendicular to 
the linear polarization direction 
({\it i.e.}, parallel to the magnetic field of the photon)
than parallel to the beam polarization.
Physically, when the magnetic interaction dominates, the 
asymmetry becomes positive, while if the electric interaction 
dominates, it becomes negative.

The DCC model calculations predict a slightly 
negative polarization over most of the photon energy range in
Fig. \ref{fig:asym}.  These results will constrain the 
backward angle predictions of theoretical models, 
which were largely unconstrained until now. 
In contrast to Ref. \cite{zegers}, where the beam asymmetry 
increases with increasing photon energy, here we see that 
$\Sigma_{K\Lambda}$ decreases with photon energy.
As shown in Fig. 11 of Ref. \cite{bruno}, where calculations 
are done without the inclusion of various $N^*$ resonances, 
the backward angle beam asymmetry is strongly affected by 
the inclusion of a third $D_{13}$ resonance (at 1954 MeV). 
The data in Fig. \ref{fig:asym} at lower photon energies 
agree better with calculations that do not include this 
third $D_{13}$ resonance. However, conclusive results can 
only be obtained by including the new data in overall fits 
to all $K\Lambda$ photoproduction data.


In summary, measurements of forward-angle $\Lambda$ production have 
been carried out at the LEPS spectrometer at SPring-8.  These data 
correspond to backward-angle kaons in the CM frame.  The data are 
in good agreement with CLAS data \cite{bradford} where the 
angular ranges overlap, and go beyond the CLAS angles 
to far backward angles.  

The cross sections agree with the 
general trend of the calculations in the DCC model \cite{bruno} 
but are significantly higher around $E_\gamma=2.1$ GeV.
The Regge model can explain the energy dependence of the present 
data, and even extrapolated to previous data at 4.3 GeV, whereas the 
effective Lagrangian models cannot reproduce the data over this
range of photon energies.
The beam asymmetries show a positive sign at photon 
energies below 2.0 GeV, in contrast with theoretical predictions
of the DCC model. 

Meson photoproduction at backward angles (and to some extent 
forward angles as well) has not been explored much so far.
Therefore, the present data are useful to constrain various 
models of strangeness production.  In the present paper, we 
have shown that coupled channels effects are important and 
that some resonance nature, with mass around 2 GeV, may be 
studied in the DCC model \cite{bruno}.
In particular, the $D_{13}$ in the $s$-channel was shown to 
affect the cross sections.
The Regge model can also provide information of hyperon 
trajectories which has not been fully studied for the 
$u$-channel.  Further experiments on kaon photoproduction 
and other mesons at backward angles will stimulate further 
theoretical progress.

 The authors thank the SPring-8 staff
 for supporting the BL33LEP beam line and the LEPS experiment. We thank
 H. Toki (RCNP) for fruitful discussions. This
 research was supported in part by the Ministry of Education, Science,
 Sports and Culture of Japan, by the National Science Council of
 Republic of China (Taiwan), Korea Research Foundation(KRF) 
 Grant(2006-312-C00507), MEC (Spain) Grant No. FIS2005-03142, 
 European Hadron Physics Project (RII3-CT-2004-506078) 
 and the National Science Foundation (NSF Award PHY-0555558).

\end{document}